\documentclass[conference]{IEEEtran}
\IEEEoverridecommandlockouts
\usepackage{tikz}
\usepackage{amsmath,amsfonts,amssymb}%
\usepackage{bm}%
\usepackage{graphicx,graphics}%
\usepackage{cases}%
\usepackage[noadjust]{cite}%
\usepackage{color}%
\usepackage{cite,url}
\usepackage{verbatim}
\usepackage{algorithm}
\usepackage{balance}
\usepackage{multirow}
\usepackage{stfloats}
\usepackage{xspace}
\usepackage{amsthm}
\usepackage{mathtools} 
\usepackage{subfig} 
\usepackage[noend]{algpseudocode}
\usepackage{caption}

\usepackage{tikz}
\usetikzlibrary{shapes.misc}
\usetikzlibrary{matrix}
\usetikzlibrary{arrows,backgrounds,fit,calc}

\makeatletter
\def\BState{\State\hskip-\ALG@thistlm}
\makeatother

\newcommand{\secref}[1]{Section\,\ref{#1}}

\DeclarePairedDelimiterX\abs[1]{\lvert}{\rvert}{#1}
\DeclarePairedDelimiterX\parn[1]{(}{)}{#1}
\DeclarePairedDelimiterX\set[1]{\lbrace}{\rbrace}{#1}
\DeclarePairedDelimiterX\innerp[2]{\langle}{\rangle}{#1,#2}
\DeclarePairedDelimiterX\norm[1]{\lVert}{\rVert}{#1}
\DeclarePairedDelimiterX\brak[1]{\lbrace}{\rbrace}{#1}
\DeclarePairedDelimiterX\coeff[1]{(}{)}{#1}

\newcommand{\matlab}{\texttt{MATLAB}}

\newcommand{\conj}[1]{\overline{#1}} 
\newcommand{\untsph}{\mathbb{S}^{2}} 

\newcommand{\lsph}{L^{2}(\mathbb{S}^{2})}

\newcommand{\bv}[1]{\boldsymbol{#1}}

\newcommand{\intsph}{\int_{\mathbb{S}^{2}}}

\newcommand{\dfn}{\triangleq}

\newcommand{\figref}[1]{Fig.\,\ref{#1}}

\newcommand{\shc}[3]{\coeff{#1}{}_{#2}^{#3}}

\graphicspath{{figs/},{Figures/}}

\newcommand{\lsphL}[1]{\mathcal{H}_{#1}}

\newcommand{\displayskipshrink}{%
    \setlength{\abovedisplayskip}{3.0pt plus 1.0pt minus 1.0pt}
    \setlength{\abovedisplayshortskip}{0pt plus 1.0pt minus 1.0pt}
    \setlength{\belowdisplayskip}{3.0pt plus 1.0pt minus 1.0pt}
    \setlength{\belowdisplayshortskip}{3.0pt plus 1.0pt minus 1.0pt}}

\begin{document}

\displayskipshrink

\title{Optimal-Dimensionality Sampling on the Sphere: Improvements and Variations}
\author{
\IEEEauthorblockN{Wajeeha Nafees\IEEEauthorrefmark{1},
Zubair Khalid\IEEEauthorrefmark{1},
Rodney A. Kennedy\IEEEauthorrefmark{2} and Jason D. McEwen\IEEEauthorrefmark{3}}

\IEEEauthorblockA{\IEEEauthorrefmark{1}School of Science and Engineering, Lahore University of Management Sciences, Lahore 54792, Pakistan}
\IEEEauthorblockA{\IEEEauthorrefmark{2}Research School of Engineering, The Australian National University, Canberra, ACT 2601, Australia }
\IEEEauthorblockA{\IEEEauthorrefmark{3}Mullard Space Science Laboratory, University College London, Surrey RH5 6NT, UK
\thanks{Rodney A. Kennedy is supported by Australian Research Council's Discovery Projects funding scheme (project no. DP150101011). Jason D. McEwen is partially supported by the Engineering and Physical Sciences Research Council (grant number EP/M011852/1).}
}

\textit{14060020@lums.edu.pk, zubair.khalid@lums.edu.pk, rodney.kennedy@anu.edu.au, jason.mcewen@ucl.ac.uk}}

\maketitle

\maketitle

\begin{abstract}
For the accurate representation and reconstruction of band-limited signals on the sphere, an optimal-dimensionality sampling scheme has been recently proposed which requires the optimal number of samples equal to the number of degrees of freedom of the signal in the spectral~(harmonic) domain. The computation of the spherical harmonic transform~(SHT) associated with the optimal-dimensionality sampling requires the inversion of a series of linear systems in an iterative manner. The stability of the inversion depends on the placement of iso-latitude rings of samples along co-latitude. In this work, we have developed a method to place these iso-latitude rings of samples with the objective of improving the well-conditioning of the linear systems involved in the computation of the SHT. We also propose a multi-pass SHT algorithm to iteratively improve the accuracy of the SHT of band-limited signals. Furthermore, we review the changes in the computational complexity and improvement in accuracy of the SHT with the embedding of the proposed methods. Through numerical experiments, we illustrate that the proposed variations and improvements in the SHT algorithm corresponding to the optimal-dimensionality sampling scheme significantly enhance the accuracy of the SHT.
\end{abstract}


\section{Introduction}
Signal analysis on spherical bodies has widespread applications in the fields of cosmology, geodesy, geomagnetics, acoustics and computer graphics~\cite{Tegmark:2004,Wieczorek:2007,Lowes:1974,Bates:2015,Zhang:2012,rama_signal:2004}. Data measured over the surface of a spherical object, i.e., in the spatial domain, can be transformed to the harmonic domain using the spherical harmonic transform~(SHT) which is the analogue in spherical geometry of the renowned Fourier transform in Euclidean geometry~\cite{Kennedy-book:2013}. Sampling schemes utilized for computing SHTs are categorized as theoretically exact, accurate or approximate~\cite{Driscoll:1994,Sneeuw:1994,Mohlenkamp:1999,Gorski:2005,Crittenden:1998,Healy:2003,Keiner:2007,Kostelec:2008,mcewen:fsht,Huffenberger:2010,McEwen:2011}. In this work, we consider those schemes which enable exact or accurate computation of the SHT of band-limited signals. Different sampling schemes have different spatial dimensionality defined as the number of sample points needed to accurately or exactly compute the SHT and thus capture the information content of band-limited signals. For the computation of SHTs of a signal band-limited at $L$~(defined in \secref{SHT-rep}), the optimal spatial dimensionality attainable by any sampling scheme on the sphere is $L^2$, which is equal to the degrees of freedom of the band-limited signal in harmonic space.

Driscoll and Healy~\cite{Driscoll:1994} developed an exact method to compute the SHT of a signal, that is band-limited at $L$, which requires $\sim~$(asymptotically, as $L\rightarrow \infty$) $4L^2$ equiangular samples on the sphere, where the complexity of most stable algorithm to compute SHT is $O(L^3)$. In comparison, the sampling scheme presented by McEwen and Wiaux~\cite{McEwen:2011} requires $\sim2L^2$ equiangular samples to exactly compute the SHT with complexity $O(L^3)$. The Gauss-Legendre sampling scheme~\cite{McEwen:2011,Skukowsky:1986} also requires $\sim2L^2$ for exact computation of the SHT, where the complexity to compute the SHT is $O(L^3)$. To the best of our knowledge, there does not exist any theoretically exact sampling scheme with dimensionality less than $\sim2L^2$. On the other hand, the SHT can also be computed approximately using the least-squares based method proposed by Sneeuw~\cite{Sneeuw:1994}, which, although requiring $L^2$ samples, becomes inaccurate and computationally inefficient scaling as $O(L^6)$ for large band-limits.

Recently, an optimal-dimensionality sampling scheme has been proposed in \cite{Khalid_OD:2014} for the accurate computation of the SHT of band-limited signals using only $L^2$ samples. Optimal-dimensionality sampling has been customized to serve the needs of applications in acoustics~\cite{Bates:2015} and diffusion MRI~\cite{Bates:2016}. Although the SHT associated with this sampling scheme requires the optimal number of samples, it has computational complexity of $O(L^{3.37})$. The computation of the SHT for optimal-dimensionality sampling involves inversion of a series of systems of linear equations. For accurate inversion of these systems, a condition number minimization method has been proposed in \cite{Khalid_OD:2014} to determine the locations of samples.

This paper aims to improve the accuracy of the SHT associated with the optimal-dimensionality sampling scheme. We serve this objective by developing a new method for the placement of samples and proposing a variation in the computation of the SHT. We develop a method, referred to as the elimination method, for the placement of iso-latitude rings of samples such that the condition number~(ratio of the largest to the smallest singular	 value) of the matrices used in the computation of the SHT is minimized. Due to the iterative nature of the resulting SHT algorithm, the error builds up in the computation of the SHT. To resolve this issue, we also propose a multi-pass SHT algorithm which iteratively reduces the residual between the given signal and the reconstructed signal. We also analyze the changes in the complexity of the SHT with the use of these methods. Through numerical experiments, we demonstrate the improvement in accuracy with the use of the proposed methods. The remainder of the paper is structured as follows. We present the necessary mathematical background in \secref{sec:prelim} before reviewing the optimal-dimensionality sampling scheme in \secref{sec:prob_formulation}. \secref{sec:developments} presents the proposed developments and also contains the accuracy analysis. 
Concluding remarks are then made in \secref{sec:conclusions}.

\section{Mathematical Background}
\label{sec:prelim}
\subsection{Signals on the Sphere}
Let $f(\theta ,\phi)$ denote a complex-valued, square integrable function on the unit sphere $\untsph$, where $\theta\in[0,\pi]$ and $\phi\in[0,2\pi)$ denote the co-latitude and longitude respectively. The space formed by these functions is a Hilbert space, denoted by $\lsph$, equipped with the following inner product given by
\begin{equation}
\label{Eq:innprd}
    \innerp{f}{h}\triangleq  \intsph f(\theta,\phi) \conj{h(\theta,\phi)}\,\sin\theta\,d\theta\,d\phi,
\end{equation}
for any two functions $f,\,h \in \lsph$. In \eqref{Eq:innprd}, $\overline{(\cdot)}$ denotes the complex conjugate operation, $\sin\theta d\theta d\phi$ is the differential surface element and $\displaystyle\intsph \equiv \displaystyle\int_{\theta=0}^{\pi} \int_{\phi=0}^{2\pi}$ is an integral over the whole sphere. The inner product in \eqref{Eq:innprd} induces a norm $\|f\| \triangleq \innerp{f}{f}^{1/2}$, and signals with finite induced norm are referred to as ``signals on the sphere".

\subsection{Harmonic Domain Representation}
\label{SHT-rep}
Signals can be transformed to the harmonic domain using the natural basis -- spherical harmonic basis functions (or simply spherical harmonics). Spherical harmonics, denoted by $Y_{\ell}^{m}(\theta, \phi)$ for integer degree $\ell\le 0 $ and integer order $-\ell\le m \le \ell$, are defined as
\begin{equation}
\label{Eq:Ylm_dfn}
Y_{\ell}^{m}(\theta, \phi)\dfn \sqrt {\frac{2\ell+1}{4\pi}\frac{(\ell-m)!}{(\ell+m)!}}\,P_\ell^m (\cos\theta)e^{im\phi},
\end{equation}
\noindent where $P_\ell^m(\cdot)$ is the associated Legendre function~\cite{Kennedy-book:2013}. Any function $f \in \lsph$ can be expanded in terms of spherical harmonics as
\begin{align}\label{Eq:f_expansion}
f(\theta,\phi)&= \sum\limits_{{\ell}=0}^{\infty}\sum\limits_{m=-{\ell}}^{\ell} \shc{f}{\ell}{m}  Y_{\ell}^{m}(\theta, \phi).
\end{align}
Here $\shc{f}{\ell}{m}$ denotes the spherical harmonic coefficient of degree $\ell\le 0 $ and order $-\ell\le m \le \ell$ and is given by the spherical harmonic transform~(SHT) as
\begin{equation} \label{Eq:flm_expansion}
 \shc{f}{\ell}{m}\dfn \langle f,Y_{{\ell}}^{m} \rangle = \intsph f(\theta,\phi) \conj{Y_\ell^m (\theta,\phi)}\,\sin\theta\,d\theta\,d\phi.
\end{equation}
The synthesis equation, \eqref{Eq:f_expansion}, to reconstruct the signal from its spherical harmonic coefficients is referred to as inverse SHT. A signal $f\in\lsph$ is said to band-limited if $\shc{f}{\ell}{m}$ = 0 for $\ell \geq L$, where $L$ is the band-limit of the signal, and can be expressed in terms of spherical harmonics as
\begin{align}\label{Eq:f_L_expansion}
f(\theta,\phi)&= \sum\limits_{{\ell}=0}^{L-1}\sum\limits_{m=-{\ell}}^{\ell} \shc{f}{\ell}{m}  Y_{\ell}^{m}(\theta, \phi).
\end{align}
The signals, band-limited at $L$, form an $L^2$ dimensional subspace of $\lsph$, which we denote by $\lsphL{L}$.

\section{Problem Formulation}
\label{sec:prob_formulation}
\subsection{Optimal Dimensionality Sampling on the Sphere}
The optimal-dimensionality sampling scheme on the sphere requires (optimal number) $L^2$ samples to accurately compute the SHT for a signal with band-limit $L$~\cite{Khalid_OD:2014}. In this scheme, $L$ iso-latitude rings are placed on the sphere at locations~(to be explained shortly) given in vector $\bv{\theta}$, defined as
\begin{equation} \label{Eq:index_theta}
\bv{\theta} \dfn \left[ \theta_0\,, \theta_1\,, \ldots ,\theta_{L-1}\right].
\end{equation}
The ring placed at $\theta_k$ contains $2k+1$ equiangular points along longitude $\phi$.

\subsection{SHT Formulation}

For a signal $f\in\lsphL{L}$ sampled using optimal-dimensionality sampling scheme, we define a vector $\bv{g}_m$, for every $|m|<L$ as
\begin{align} \label{Eq:g_m_dfn}
\bv{g}_m \dfn \ \left[G_m(\theta_{|m|}) , G_m(\theta_{|m|+1}), \ldots , G_m(\theta_{L-1}) \right]^T,
\end{align}
where $G_m(\theta_k)$ for each $\theta_k \in \bv{\theta}$ is given as
\begin{align} \label{Eq:G_m_theta_k}
G_m(\theta_{k}) \dfn \int_{0}^{2\pi} f(\theta_k,\phi) e^{-im\phi} d\phi &= 2\pi \sum_{\ell=m}^{L-1} \shc{f}{\ell}{m} \tilde{P}_\ell^m (\theta_k).
\end{align}
\noindent Here $\tilde{P}_\ell^m (\theta_k) \dfn Y_{\ell}^{m}(\theta_k,0)$ denotes scaled associated Legendre functions. The second equality in \eqref{Eq:G_m_theta_k} is obtained by using \eqref{Eq:Ylm_dfn} and \eqref{Eq:f_L_expansion} and employing the orthogonality of complex exponentials.
By defining another vector $\bv{f}_m$ as
\begin{align} \label{Eq:f_m_dfn}
\bv{f}_m=\left[\shc{f}{|m|}{m}, \shc{f}{|m|+1}{m}, \ldots,\shc{f}{L-1}{m} \right]^T,
\end{align}
containing the spherical harmonic coefficients of order $m$, we formulate a linear system given as
\begin{align} \label{Eq:system_eq}
\bv{g}_m=\bv{P}_m\,\bv{f}_m,
\end{align}
where the $\bv{P}_m$ is an $(L-|m|)\times (L-|m|)$ matrix with elements given by
\begin{align} \label{Eq:Pm_dfn}
\bv{P}_m(i,j)=\tilde{P}_{|m|+j-1}^m (\theta_{|m|+i-1}).
\end{align}

\subsection{Problem Under Consideration}

For each order $|m|\le L$, the spherical harmonic coefficients contained in $\bv{f}_m$ can be recovered by solving the linear system given in \eqref{Eq:system_eq}. Computation of the SHT, i.e., the computation of spherical harmonic coefficients of the signal $f\in\lsphL{L}$ sampled according to the optimal-dimensionality sampling scheme, involves the inversion of a series of linear systems formed by the matrix $\bv{P}_m$~(defined in \eqref{Eq:Pm_dfn}) for $m=0,1,\hdots,L-1$~\cite{Khalid_OD:2014}. A condition number minimization method has been proposed in \cite{Khalid_OD:2014} to determine the locations of the iso-latitude rings indexed in \eqref{Eq:index_theta} such that the matrix $\bv{P}_m$ for each $m=0,1,\hdots,L-1$ is well-conditioned and the SHT can be accurately computed. With an objective to improve the accuracy of the SHT, we consider the problem of determining the locations of iso-latitude rings of samples which reduce~(improve) the condition number~(ratio of the largest to the smallest eigenvalue) of the matrices $\bv{P}_m,\,\, m=0,1,\hdots,L-1$. The $\bv{\theta}$ vector containing the locations of iso-latitude rings, initially in the ascending order of co-latitude angle, is re-ordered such that every $\bv{P}_m$ matrix has minimum condition number. To further increase the accuracy of the SHT, we also propose a multi-pass SHT algorithm which iteratively reduces the error between the given signal~(samples in spatial domain) and the signal synthesized using the computed spherical harmonic coefficients.

\section{Optimized Samples Placement and Multi-Pass SHT}
\label{sec:developments}
With an objective to improve the accuracy of the SHT using optimal number of samples, we first present our proposed method for the placement of the iso-latitude rings of samples and later we present iterative method for the computation of the SHT.

\subsection{Condition Number Minimization}

The recovery of $\bv{f}_m$ for each $|m|<L$ using \eqref{Eq:system_eq} requires inversion of the $\bv{P}_m$ matrix for each $|m|<L$. For accurate computation of the SHT, it is therefore necessary that the matrix $\bv{P}_m$ is invertible and well-conditioned. Since $\bv{P}_m$ is a matrix of associated Legendre polynomials of order $m$ and degrees $|m| \le \ell <L $ evaluated at $\theta_i,\,\, i=|m|, |m+1|,\hdots, L-1$, its accurate inversion depends on the locations of the iso-latitude rings indexed in \eqref{Eq:index_theta}. To determine the locations of the iso-latitude rings, we propose a condition number minimization technique, herein referred to as the \emph{elimination method}, for the construction of the vector $\bv{\theta}$.

Let $\Omega$ be a set of $L$ equiangular co-latitude angles between $0$ and $\pi$ defined as
\begin{equation} \label{Eq:set_omega}
\Omega \dfn \left\{\frac{\pi\,(2t+1)}{2L-1}\right\}\,, \quad t = 0,1,\ldots, L-1.
\end{equation}
\noindent For $m=0$, the $\bv{P}_{m}$ matrix is formed by inserting all elements of set $\Omega$ in \eqref{Eq:Pm_dfn} and has dimension $L\times L$. Since $\bv{P}_{m}$, for $m$=1, requires $L-1$ co-latitude angles, we eliminate one element, say $\left\{\Omega_j\right\}$, from the set $\Omega$ and calculate the condition number, denoted by $\kappa_m$, of $\bv{P}_{m}$ using all possible $L-1$ combinations of residual elements $\Omega \backslash \left\{\Omega_j\right\}$. The element $\left\{\Omega_j\right\}$, whose elimination results in the lowest condition number of $\bv{P}_m$, is then selected as the first element of the $\bv{\theta}$ vector. The $\Omega$ set is then updated as $\Omega\leftarrow\Omega\backslash \left\{\Omega_j\right\}$. The same procedure is carried out for the construction of the $\bv{\theta}$ vector for $m=2,3,\ldots,L-1$ which we summarize below in the form of an algorithm.

\begin{algorithm}
\caption{Elimination Method}\label{min_cond}
\begin{algorithmic}[1]
\Require $\bv{\theta}$ given $L$
\Procedure{Elimination method}{}
\State $\Omega \dfn \left\{\frac{\pi\,(2t+1)}{2L-1}\right\}_{t = 0,1,\ldots, L-1}.$
\For{$m = 0, 1, \ldots, L-1$}
					\For{$j = 0, 1, \ldots, L-m$}
									\State $\alpha_j \leftarrow \Omega \backslash \left\{\Omega_j\right\}$
									\State evaluate $\bv{P}_m$ using \eqref{Eq:Pm_dfn} and
									\State compute condition number $\kappa_m$
									\EndFor
					\State \textbf{end for}
			\State determine index $k$ for minimum value of $\kappa_m$
			\State update $\theta_m \leftarrow \Omega_k $
			\State update $\Omega \leftarrow \alpha_k $
			\EndFor
\State \textbf{end for}
			\State \textbf{return} $\bv{\theta}$
\EndProcedure
\State \textbf{end procedure}
\end{algorithmic}
\end{algorithm}

\begin{figure}[!ht]
\includegraphics[width=8.7cm, height=6.65cm]{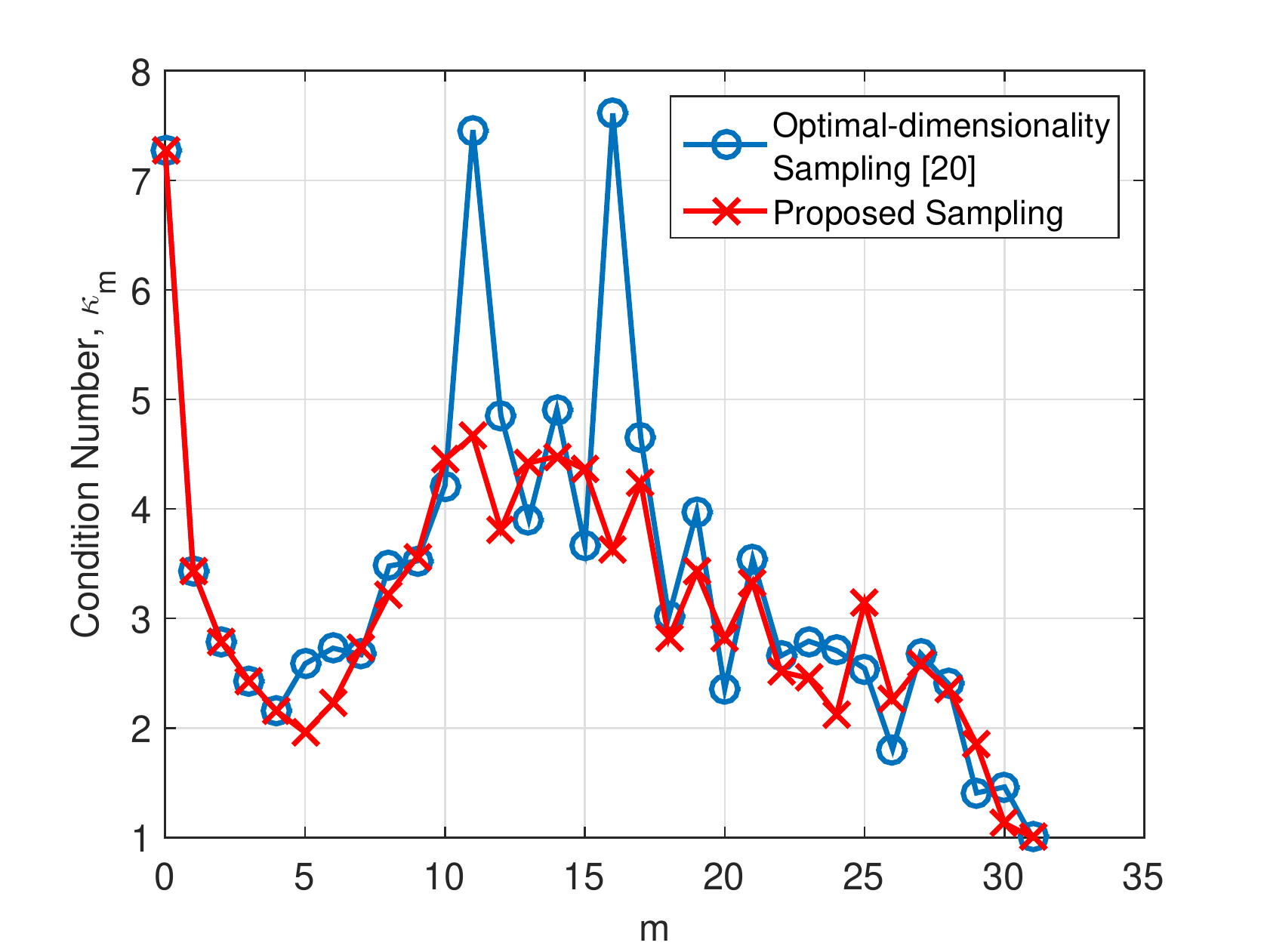}
\caption{The condition number $\kappa_m$ of the matrix $\bv{P}_m, \,\,m=0,1,\hdots,L-1$ using the proposed optimized placement of iso-latitude rings and the design proposed in \cite{Khalid_OD:2014} for band-limit $L=32$.
}
\label{fig:min_cond_no}
\centering
\end{figure}

\begin{figure}[!ht]
\includegraphics[width=8.7cm, height=6.65cm]{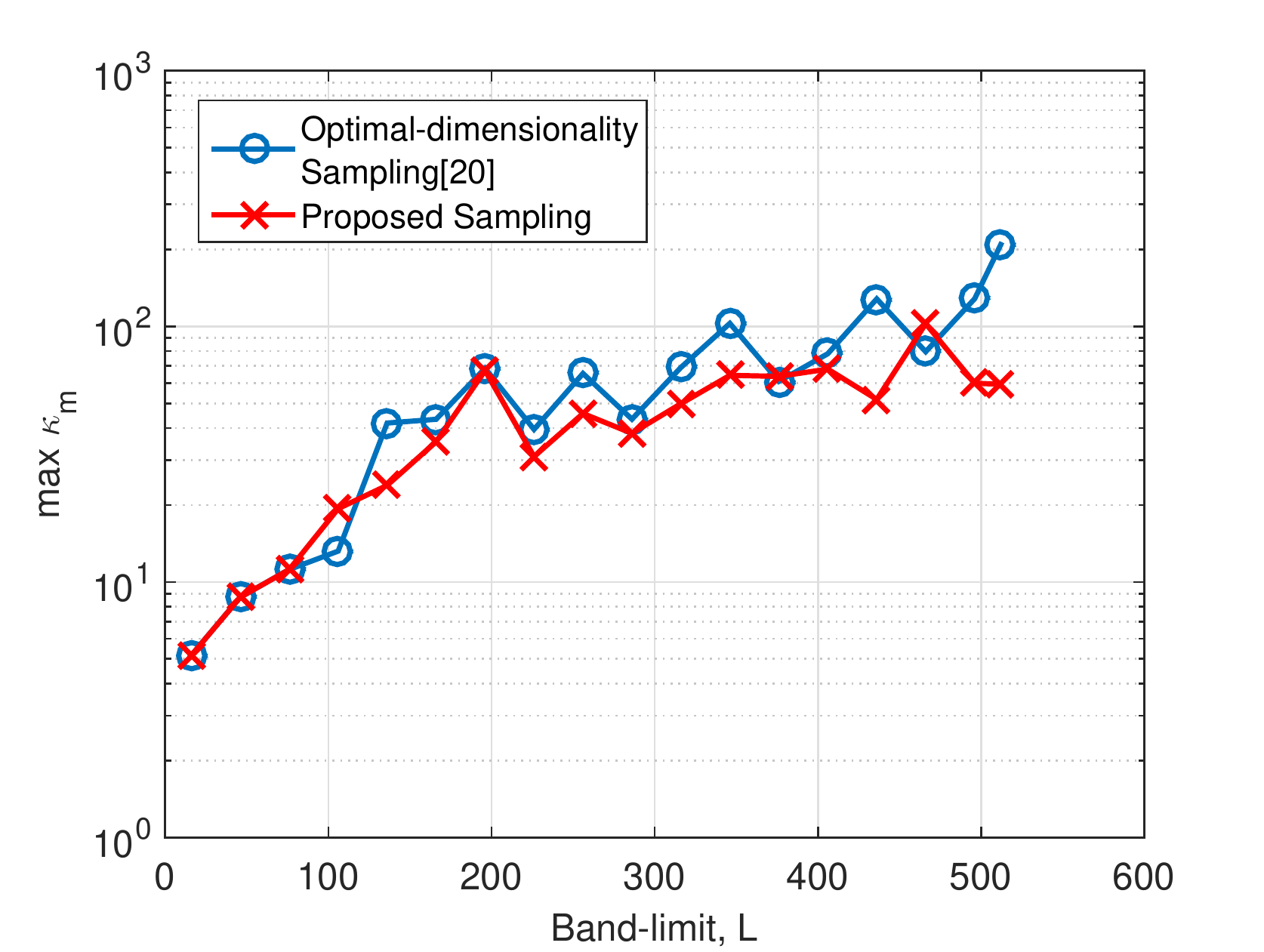}
\caption{The maximum of the condition number $\max(\kappa_m)\,,\,\, 0\leq m <L$ for different band-limits $16 \leq L\leq 512$. }
\centering
\label{fig:max_cond2}
\end{figure}

The $\bv{\theta}$ vector constructed using the proposed elimination method is optimized in a sense that it generates $\bv{P}_m$ matrices of lower condition number as compared to the optimal-dimensionality sampling scheme. This improvement in the condition number comes from the fact that the proposed elimination method has $L-|m|$ choices for $\theta_m$ such that the condition number of matrix $\bv{P}_m$ is minimized. In contrast, the method proposed in \cite{Khalid_OD:2014} has $|m|$ choices for the selection of $\theta_m$ and minimization of the condition number of matrix the $\bv{P}_m$. As an illustration, the condition number $\kappa_m$ of the matrix $\bv{P}_m, \,\,m=0,1,\hdots,L-1$ using the proposed optimized placement of iso-latitude rings and the design proposed in \cite{Khalid_OD:2014} is plotted in \figref{fig:min_cond_no} for band-limit $L=32$. We also plot the maximum of the condition number $\kappa_m$ obtained for different band-limits $16 \leq L \leq 512$ in \figref{fig:max_cond2}. It is evident that the proposed elimination method improves the well-conditioning of the systems involved in the computation of the SHT algorithm associated with the optimal-dimensionality sampling on the sphere.

\subsection{Multi-pass SHT}
The spherical harmonic coefficients of a band-limited signal sampled according to the optimal-dimensionality sampling scheme are computed iteratively for each order in a sequence $|m| = L-1, L-2, \hdots, 0$.
The SHT is inherently iterative in nature as the spherical harmonic coefficients of order $|m|$ are used in the computation of the SHT of order $|m|-1$. Consequently, the error propagates and builds up in the iterative computation of spherical harmonic coefficients. To reduce this error building-up, we propose a multi-pass SHT algorithm which iteratively improves the accuracy of the SHT.

For a signal $f\in\lsphL{L}$ sampled by the optimal-dimensionality sampling scheme, the spherical harmonic coefficients can be accurately computed by the algorithm presented in \cite{Khalid_OD:2014}\footnote{SHT can be computed accurately for band-limited signals sampled over optimal-dimensionality sampling scheme~\cite{Khalid_OD:2014} using the~\matlab~ based package Novel Spherical Harmonic Transform~(NSHT) publicly available at \url{www.zubairkhalid.org/nsht}.}. We define the residual~(error) between the signal $f$ and the signal synthesized from the recovered spherical harmonic coefficients as
\begin{align}
\label{Eq:residual}
r_k(\theta,\phi) = f(\theta,\phi) - \sum_{\ell=0}^{L-1}\sum_{m=-\ell}^{\ell} \shc{\tilde{f_k}}{\ell}{m}  Y_\ell^m(\theta,\phi)
\end{align}
where $\shc{\tilde{f_k}}{\ell}{m} $ denotes the spherical harmonic coefficient computed using the proposed SHT algorithm and $k=1$~(indicating the number of times the transform has been carried out). Once residual is computed, we use the SHT algorithm to compute its spherical harmonic coefficients, denoted by $\shc{\tilde{r_k}}{\ell}{m}$, which we use to update $\shc{\tilde{f_k}}{\ell}{m}$ as
\begin{align}
\label{Eq:update}
\shc{\tilde{f_{k+1}}}{\ell}{m} = \shc{\tilde{f_k}}{\ell}{m} + \shc{\tilde{r_k}}{\ell}{m}.
\end{align}
We propose to iteratively use \eqref{Eq:residual} and \eqref{Eq:update} to compute $\shc{\tilde{f_{k}}}{\ell}{m}$ for $k=1,2,\hdots$, until the following stopping criterion is met
\begin{align}
\label{Eq:criterion}
\max |r_{k+1}(\theta,\phi)| > \max |r_{k}(\theta,\phi)|,
\end{align}
where $\max$ is taken over the samples of the sampling scheme. Since the proposed method requires to compute the SHT multiple times, we refer to the proposed method for the computation of spherical harmonic coefficients as the multi-pass SHT. Later, we illustrate that the proposed method significantly improves the accuracy of the SHT.

\subsection{Computational Complexity Analysis}
Here we briefly discuss the computational complexity of the proposed elimination method for the placement of iso-latitude rings and the multi-pass SHT algorithm. The elimination method has the computational complexity of $O(L^5)$. However, it only needs to be run once for the determination of $\bv{\theta}$ for each $L$. Furthermore, we note that the complexity of the method presented in \cite{Khalid_OD:2014} for the placement of samples is also $O(L^5)$. For the optimal-dimensionality sampling scheme, the SHT can be computed with complexity of $O(L^{3.37})$. For the proposed multi-pass SHT algorithm, the complexity scales with the number of iterations, denoted by $K$, needed for the convergence of error. In the next section, we provide examples to illustrate that the proposed multi-pass SHT algorithm converges quickly in $K\ll L$ number of iterations.

\subsection{Accuracy Analysis}

\begin{figure}
\includegraphics[width=8.7cm, height=6.65cm]{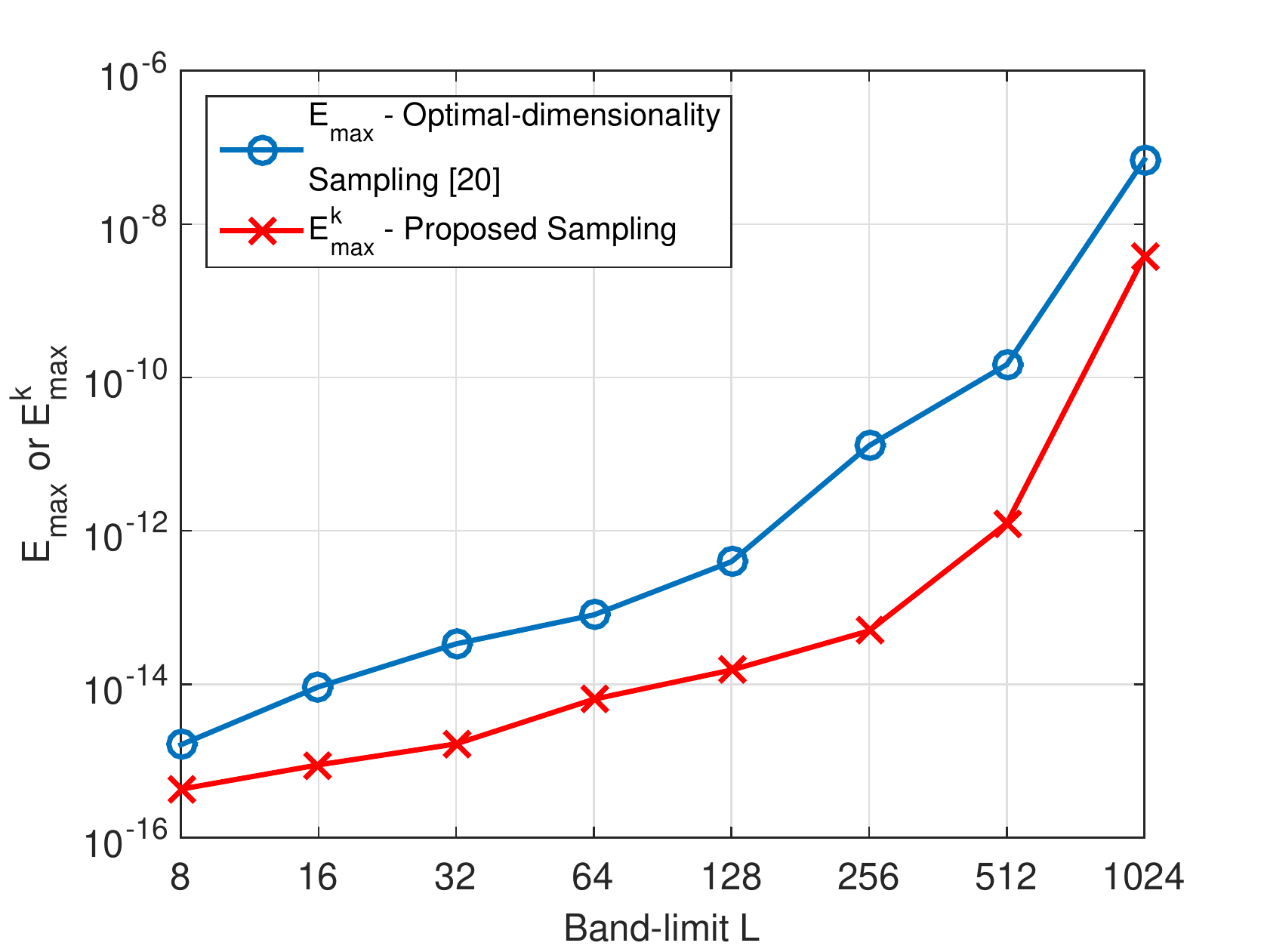}
\caption{Maximum errors $E_{\max}$ and $E_{\max}^k$ between the original and recovered spherical harmonic coefficients using original optimal-dimensionality sampling scheme and the proposed sampling respectively for band-limits $8\le L\le1024$. Here $k$ depends on the stopping criterion given in \eqref{Eq:criterion} and is different for each band-limit $L$.
}
\label{fig:accuracy}
\centering
\end{figure}

In this section, we analyse the accuracy of the proposed multi-pass SHT algorithm of a band-limited signal evaluated using the optimal-dimensionality sampling scheme with iso-latitude rings placed using the proposed elimination method. 
Comparison between the proposed developments and the SHT proposed in \cite{Khalid_OD:2014} has been carried out through numerical experiments. 
In our experiment, we first take a band-limited signal $f\in\lsphL{L}$ by randomly generating its spherical harmonic coefficients $\shc{f}{\ell}{m}$. The real and imaginary parts of the coefficients are uniformly distributed in $[0,1]$. 
Using inverse SHT, we obtain the signal $f$ in the spatial domain, that is, over the samples of the optimal-dimensionality sampling scheme~(proposed sampling or \cite{Khalid_OD:2014}).
We then apply the SHT presented in \cite{Khalid_OD:2014} and the proposed multi-pass SHT algorithm to recover the spherical harmonic coefficients, denoted by $\shc{\tilde{f}}{\ell}{m}$ and $\shc{\tilde{f_k}}{\ell}{m}$ respectively. We conduct experiments for 10 different signals to obtain the average value of the maximum error between reconstructed and original spherical harmonic coefficients defined as
\begin{align} \label{Eq:max_error}
E_{\max} &\dfn \max |\shc{\tilde{f}}{\ell}{m}-\shc{f}{\ell}{m}|, \\
\label{Eq:max_error_2}
E_{\max}^k &\dfn \max |\shc{\tilde{f}_k}{\ell}{m}-\shc{f}{\ell}{m}|,
\end{align}
which we plot for band-limits $8\le L\le1024$ in \figref{fig:accuracy}, where it can be observed that the proposed multi-pass SHT algorithm and optimized placement of samples results in the more accurate computation of the SHT.

\begin{figure}
\includegraphics[width=8.7cm, height=6.65cm]{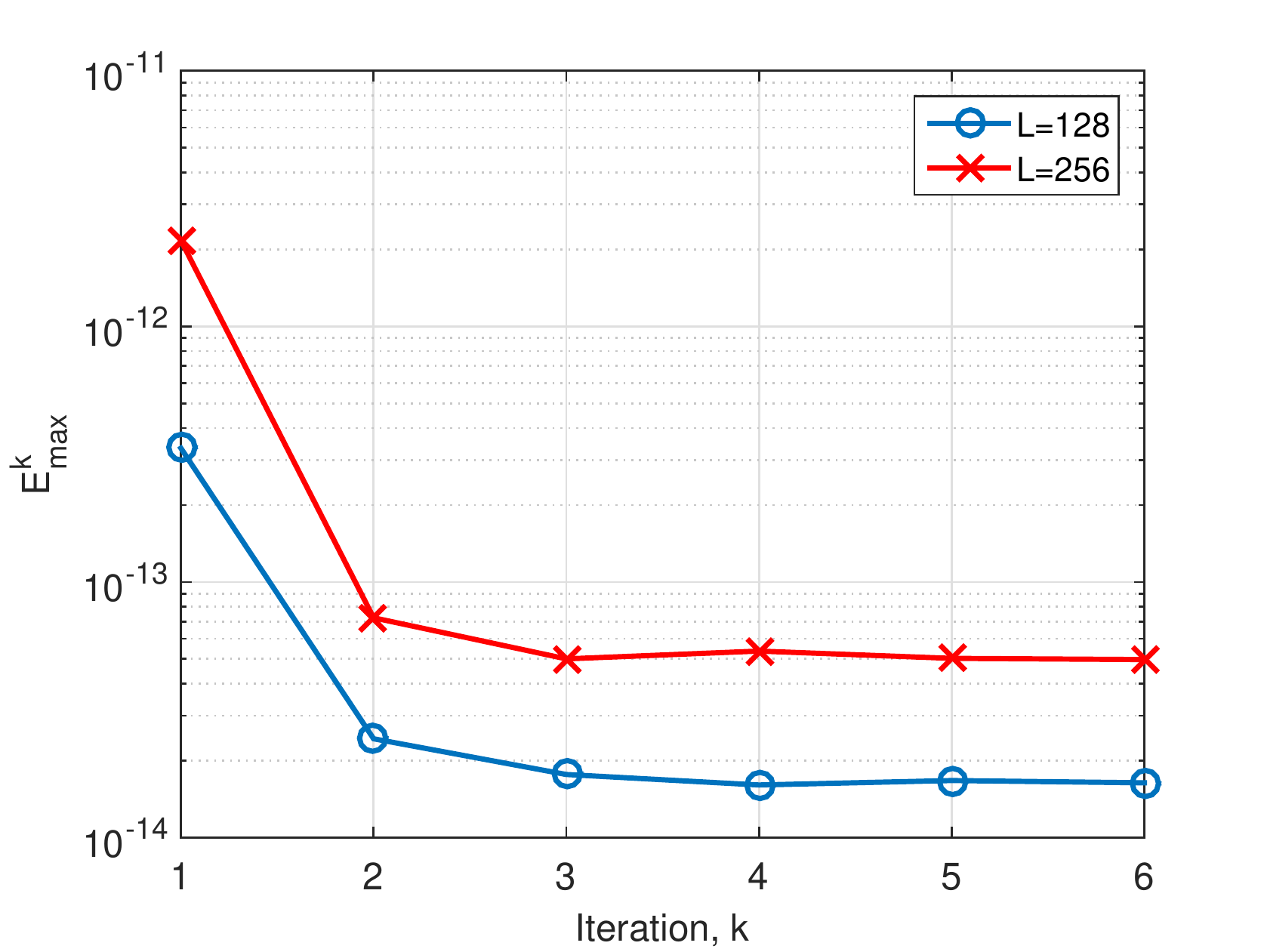}
\caption{Maximum error $E_{\max}^k$, given in \eqref{Eq:max_error_2}, between the original and recovered spherical harmonic coefficients for band-limits $L=128$ and $L=256$ and different iterations of the multi-pass SHT.}
\label{fig:multipass}
\centering
\end{figure}

We also analyse the convergence of the multi-pass SHT algorithm and the improvement in the accuracy of the SHT enabled by the proposed multi-pass SHT algorithm. We plot the maximum absolute error $E_{\max}^k$ for band-limits $L=128$ and $L=256$ in \figref{fig:multipass}, where it can be observed that the proposed multi-pass SHT improves the accuracy of SHT and converges~(quickly) in $K\ll L$ number of iterations.

\section{Conclusions}
\label{sec:conclusions}

In this work, we have proposed variations in the spherical harmonic transform~(SHT) associated with the optimal-dimensionality sampling scheme which consist of iso-latitude rings of samples and enables accurate computation of the SHT of band-limited signals using the optimal number of samples given by the degrees of freedom required to represent a band-limited signal in harmonic space. We have presented the elimination method for the iterative placement of iso-latitude rings of samples. The proposed placement reduces the condition number of matrices involved in the computation of SHT and consequently improves the accuracy of the SHT. We have also proposed the multi-pass SHT algorithm which iteratively reduces the residual between the given signal and the reconstructed signal and therefore improves the overall accuracy of the SHT. We have analyzed the changes in the computational complexity and improvement in accuracy with the use of proposed variations in the computation of the SHT. We have also conducted numerical experiment to illustrate the improvement in accuracy enabled by the proposed methods.

\bibliography{IEEEabrv,sht_bib}

\end{document}